%% file: main.tex
\documentclass{article}

\input{header}

\title{Two-stage Risk Control with Application to Ranked Retrieval}
 \input{files/authors}

\date{\today}

\begin{document}
\maketitle


\input{files/abstract}
 \input{files/introduction}
 \input{files/setup}
 \input{files/method}

\input{files/application}

\input{files/conclusion}

 \bibliographystyle{plainnat}
 \bibliography{ref} 

 \newpage

\input{files/appendix}

\end{document}

%% file: header.tex
\usepackage{epsfig}
\usepackage[round]{natbib}
\usepackage{amsmath}
\usepackage{color}
\usepackage{booktabs}
\usepackage{amssymb}
\usepackage{subfig}

\usepackage{authblk}

\usepackage[colorlinks=true,citecolor=blue,urlcolor=blue,linkcolor=blue]{hyperref}
\usepackage{algpseudocode}

\def\bSig\mathbf{\Sigma}

\newcommand\BibTeX{{\rmfamily B\kern-.05em \textsc{i\kern-.025em b}\kern-.08em
T\kern-.1667em\lower.7ex\hbox{E}\kern-.125emX}}

\usepackage[dvips, letterpaper, nohead, top=1.3in, bottom=1.3in, left=1.3in, right=1.3in]{geometry}

\input{files/macro}

\newtheorem{theos}{Theorem}

\newtheorem{cors}{Corollary}

\newtheorem{ass}{Assumption}

%% file: files/macro.tex
\newcommand{\Exs}{\ensuremath{{\mathbb{E}}}}
\newcommand{\Prob}{\ensuremath{{\mathbb{P}}}}

\DeclareMathOperator{\argmin}{argmin}

\newcommand{\Xspace}{\ensuremath{\mathcal{X}}}
\newcommand{\Yspace}{\ensuremath{\mathcal{Y}}}
\newcommand{\Zspace}{\ensuremath{\mathcal{Z}}}

\newcommand{\Data}{\mathcal{D}}

\newcommand{\R}{\mathbb{R}}

\newcommand{\indic}{\mathbf{1}}

\newcommand{\lb}{\left\{ }
\newcommand{\rb}{\right \}}

\newcommand{\lp}{\left (}
\newcommand{\rp}{\right )}

\newcommand{\fL}{L^{\scalebox{0.6}{(1)}}_i}
\newcommand{\sL}{L^{\scalebox{0.6}{(2)}}_i}

\newcommand{\fll}{l^{\scalebox{0.6}{(1)}}}
\newcommand{\sll}{l^{\scalebox{0.6}{(2)}}}

\newcommand{\fR}{R^{\scalebox{0.6}{(1)}}}
\newcommand{\sR}{R^{\scalebox{0.6}{(2)}}}
\newcommand{\feR}{\widehat{R}_n^{\scalebox{0.6}{(1)}}}
\newcommand{\seR}{\widehat{R}_n^{\scalebox{0.6}{(2)}}}

\newcommand{\feRi}{\widehat{R}_n^{\scalebox{0.6}{(1)}}(\lambda_i)}
\newcommand{\seRij}{\widehat{R}_n^{\scalebox{0.6}{(2)}}(\lambda_i, \gamma_j)}

\newcommand{\fLi}[1]{L^{\scalebox{0.6}{(1)}}_i(#1)}
\newcommand{\fLn}[1]{L^{\scalebox{0.6}{(1)}}_{n+1}(#1)}
\newcommand{\sLi}[1]{L^{\scalebox{0.6}{(2)}}_i(#1)}
\newcommand{\sLn}[1]{L^{\scalebox{0.6}{(2)}}_{n+1}(#1)}

\newcommand{\tfLi}[1]{\tilde{L}^{\scalebox{0.6}{(1)}}_i(#1)}

\newcommand{\tsLi}[1]{\tilde{L}^{\scalebox{0.6}{(2)}}_i(#1)}

\newcommand{\lzone}{\lambda_{0}^{\scalebox{0.6}{(1)}}}
\newcommand{\hlzone}{\hat{\lambda}_{0}^{\scalebox{0.6}{(1)}}}

\newcommand{\lztwo}{\lambda_{0}^{\scalebox{0.6}{(2)}}}
\newcommand{\hlztwo}{\hat{\lambda}_{0}^{\scalebox{0.6}{(2)}}}

\newcommand{\gztwo}{\gamma_{0}^{\scalebox{0.6}{(2)}}}
\newcommand{\hgztwo}{\hat{\gamma}_{0}^{\scalebox{0.6}{(2)}}}

\newcommand{\hlonet}{\hat{\lambda}^{\scalebox{0.6}{(1)}}(t)}
\newcommand{\hlone}{\hat{\lambda}^{\scalebox{0.6}{(1)}}}
\newcommand{\lonet}{\lambda^{\scalebox{0.6}{(1)}}(t)}

\newcommand{\hgtwot}{\hat{\gamma}^{\scalebox{0.6}{(2)}}(t)}

\newcommand{\Hione}{\mathcal{H}^{\scalebox{0.6}{(1)}}_i}
\newcommand{\Hone}{\mathcal{H}^{\scalebox{0.6}{(1)}}}
\newcommand{\Hipone}{\mathcal{H^\prime}^{\scalebox{0.6}{(1)}}_i}

\newcommand{\Hijtwo}{\mathcal{H}^{\scalebox{0.6}{(2)}}_{i,j}}
\newcommand{\Hijptwo}{\mathcal{H^\prime}^{\scalebox{0.6}{(2)}}_{i,j}}
\newcommand{\Htwo}{\mathcal{H}^{\scalebox{0.6}{(2)}}}

\newcommand{\pione}{p_i^{\scalebox{0.6}{(1)}}}
\newcommand{\pone}{p^{\scalebox{0.6}{(1)}}}
\newcommand{\pijtwo}{p_{i,j}^{\scalebox{0.6}{(2)}}}
\newcommand{\ptwo}{p^{\scalebox{0.6}{(2)}}}

\newcommand{\Fone}{\mathcal{F}^{\scalebox{0.6}{(1)}}}
\newcommand{\Fitwo}{\mathcal{F}_i^{\scalebox{0.6}{(2)}}}
\newcommand{\Ftwo}{\mathcal{F}^{\scalebox{0.6}{(2)}}}

\newcommand{\Rejone}{\mathcal{R}^{\scalebox{0.6}{(1)}}}
\newcommand{\Rejtwo}{\mathcal{R}^{\scalebox{0.6}{(2)}}}

\newcommand{\fConf}[1]{\hat{\mathcal{C}}^{\scalebox{0.6}{(1)}}(#1)}

\newcommand{\sConf}[1]{\hat{\mathcal{C}}^{\scalebox{0.6}{(2)}}(#1)}

\newcommand{\be}{\begin{equation}}
\newcommand{\ee}{\end{equation}}

\newcommand{\ben}{\begin{equation*}}
\newcommand{\een}{\end{equation*}}

\newcommand{\mretrieval}{\mathcal{M}_{\text{retrieval}}}
\newcommand{\mrank}{\mathcal{M}_{\text{rank}}}

\newcommand{\dataone}{\mathcal{I}_1}
\newcommand{\datatwo}{\mathcal{I}_2}

\newcommand{\tlzone}{\tilde{\lambda}_{0}^{\scalebox{0.6}{(1)}}}
\newcommand{\tlonet}{\tilde{\lambda}^{\scalebox{0.6}{(1)}}(t)}
\newcommand{\tlone}{\tilde{\lambda}^{\scalebox{0.6}{(1)}}}

\newcommand{\tgztwo}{\tilde{\gamma}_{0}^{\scalebox{0.6}{(2)}}}
\newcommand{\tgtwot}{\tilde{\gamma}^{\scalebox{0.6}{(2)}}(t)}

\newcommand{\bgtwo}{\bar{\gamma}^{\scalebox{0.6}{(2)}}}

\newcommand{\levelk}{\alpha_k}

\newcommand{\levelone}{\alpha_1}
\newcommand{\leveltwo}{\alpha_2}

\newcommand{\rejset}{\mathcal{R}}

%% file: files/authors.tex
\author[1]{Yunpeng Xu\thanks{These authors contributed equally.}}
\author[2]{Mufang Ying$^*$}
\author[3]{Wenge Guo\thanks{Author e-mail addresses: yx8@njit.edu, my426@scarletmail.rutgers.edu, wenge.guo@njit.edu, zhi.wei@njit.edu}}
\author[1]{Zhi Wei}
\affil[1]{Department of Computer Science\\
       New Jersey Institute of Technology}
\affil[2]{Department of Statistics\\ Rutgers University - New Brunswick}
\affil[3]{Department of Mathematical Sciences\\
       New Jersey Institute of Technology}

%% file: files/abstract.tex
\begin{abstract}
Practical machine learning systems often operate in multiple sequential stages, as seen in ranking and recommendation systems, which  typically include a retrieval phase followed by a ranking phase. Effectively assessing prediction uncertainty and ensuring effective risk control in such systems pose significant challenges due to their inherent complexity. 
To address these challenges, we developed two-stage risk control methods based on the recently proposed learn-then-test (LTT) and conformal risk control (CRC) frameworks. Unlike the methods in prior work that address multiple risks, our approach leverages the sequential nature of the problem, resulting in reduced computational burden. We provide theoretical guarantees for our proposed methods and design novel loss functions tailored for ranked retrieval tasks. The effectiveness of our approach is validated through experiments on two large-scale, widely-used datasets: MSLR-Web and Yahoo LTRC.

\vskip 10pt

\end{abstract}

%% file: files/introduction.tex
\section{Introduction}
\label{intro}
As machine learning models become more integrated into our daily lives, the need for transparency and reliability in their predictions is crucial. Moving beyond black-box approaches, the ability to understand and quantify uncertainty in these models is increasingly important to ensure their effectiveness in real-world applications. Conformal prediction, a distribution-free and statistically valid approach that is straightforward to integrate with existing models, has emerged as a promising solution for quantifying uncertainty in machine learning \citep{vovk2005}. Its primary objective is to generate uncertainty sets for model predictions while ensuring a specified coverage level. Recently, the conformal risk control framework \citep{angelopoulos2022conformal} has expanded upon traditional miscoverage control by enabling control over the expected value of any loss function. This extension greatly enhances its applicability across a wider range of contexts.

Most existing research on conformal prediction focuses on a single-stage process, where the machine learning system processes the input and generates the prediction in a single step. However, this assumption does not hold for many real-world systems, which often involve two or more concatenated stages. One notable example is ranked retrieval systems, such as search engines, where the task is to retrieve and rank documents based on their relevance to a user's query.
These systems typically involve two sequential stages: (1) the retrieval stage, which identifies a set of candidate documents from a large repository, and (2) the ranking stage, which refines and orders these candidates to produce the final ranked list presented to the user \citep{dyin_kdd16, okhattab_sigir20}. This two-stage approach is necessary because the massive volume of documents often exceeds the capacity of a single-stage ranking model, particularly when employing computationally intensive methods. In such two-stage problems, each stage is designed with distinct optimization objectives, and errors from one stage can propagate to the next. Consequently, the two-stage process introduces additional complexity, making it more challenging to accurately quantify and control uncertainty. 

To address these challenges, we propose \textit{two-stage} conformal prediction methods to quantify and control the uncertainty inherent in such problems. Specifically, we apply the learn-then-test framework  \citep{angelopoulos2021learn} and extend the recently developed single-stage conformal risk control framework \citep{angelopoulos2022conformal}  to a two-stage setup, where each stage has its own distinct risk control requirement. Risk control is achieved by identifying parameters that  jointly satisfy the risk constraints for both stages. Furthermore, to address the specific purpose of the two stages in ranked retrieval problems, we introduce the \textit{retrieval risk} and the \textit{ranking risk}, respectively, and then apply our proposed two-stage risk control methods to derive their corresponding prediction sets while controlling both risks at the pre-specified levels. Our proposed methods are model-agnostic and can be seamlessly integrated into existing ranked retrieval systems.

\subsection*{Related work}

\paragraph{Conformal prediction}  Conformal prediction, originally developed by Vovk and collaborators, has recently emerged as a prominent method for uncertainty quantification in statistical machine learning \citep{Vovk1999, papadopoulos2002, vovk2005,lei2015}. A recent survey by Angelopoulos and Bates outlines the significance and wide applications of the topic \citep{Angelopoulos2021}. Our work builds upon the learn-then-test framework \citep{angelopoulos2021learn} and the recently developed conformal risk control (CRC) framework \citep{angelopoulos2022conformal}. The application discussed in our work shares similarities with \citep{Angelopoulos23}, which uses the LTT technique \citep{angelopoulos2021learn} to control the false discovery rate in recommender systems and optimize recommendation diversity. However, despite addressing the same ranked retrieval challenges, our work and that of \citep{Angelopoulos23} differ in both objectives and methodologies. 

\paragraph{Ranked retrieval} Ranked retrieval has been extensively studied, with models evolving from traditional IR approaches like BM25 \citep{rbyates_99, robertson76} to modern learning-to-rank algorithms \citep{Liu2009LearningTR}. Recent advances in deep learning have also been successfully applied to ranked retrieval \citep{aseveryn_sigir15, jguo_sigir16}, with methods broadly categorized into pointwise \citep{NIPS2001_5531a583, wchu_icml05}, pairwise \citep{cburges_icml05, yfreund_jmlr03}, and listwise approaches \citep{CBurges2006, zcao_msr_tr}, based on their loss functions. A recent study~\citep{wang2023uncertainty} also examines a similar two-stage problem in recommender systems but focuses on group fairness in the first stage, differing from our objective. Another work~\citep{guo2023inference} introduces stochastic ranking at inference to ensure utility or fairness in learning-to-rank models. In contrast, our work addresses a distinct focus.

%% file: files/setup.tex
\section{Problem setup}
\label{sec:setup}
Formally, in the first stage, consider an i.i.d collection of  non-increasing, right-continuous random functions $\fL: \Lambda \to [0,1]$, $i = 1, \ldots, n+1$, representing the associated losses. We denote by $\lambda$ the tuning parameter in this stage. In the second stage, we consider another i.i.d collection of random functions, $\sL: \Lambda \times \Gamma \to [0,1]$, $i = 1, \ldots, n+1$, to represent the associated losses in the second stage, incorporating an additional tuning parameter $\gamma$. Here, $\sLi{\lambda, \gamma}$ is assumed to be non-increasing and right continuous in each coordinate. Furthermore, the following conditions hold: $\fLi{0} = 1$, $\fLi{1} = 0$, $\sLi{0,0} = 1$, and $\sLi{1,1} = 0$.  We use $\fR(\lambda)$ and $\sR(\lambda, \gamma)$ to denote the expected risk functions at each stage with fixed tuning parameter $\lambda$ and $\gamma$, i.e., $\fR(\lambda) = \Exs \fLn{\lambda}$ and $\sR(\lambda, \gamma) = \Exs \sLn{\lambda, \gamma}$, and use $\feR(\lambda)$ and $\seR(\lambda, \gamma)$ to denote the empirical risk functions, i.e., $\feR(\lambda) = \frac{1}{n} \sum_{i = 1}^n \fLi{\lambda}$ and $\seR(\lambda, \gamma) = \frac{1}{n}\sum_{i = 1}^n \sLi{\lambda, \gamma}$. Without loss of generality, we assume that the parameter $\lambda$ is chosen from a  finite set $\Lambda=\left\{\lambda_i: i \in [m]\right\}$, and the parameter $\gamma$ is chosen from a finite set $\Gamma=\left\{\gamma_i: i\in[m]\right\}$. We assume the values are ordered such that $0 \leq \lambda_1<$ $\lambda_2<\cdots<\lambda_m \leq 1$ and $0 \leq \gamma_1<\gamma_2<\cdots<\gamma_m \leq 1$.

For pre-specified risk levels $\levelone,\leveltwo$, we aim to utilize random functions $\{\fLi{\lambda}\}_{i = 1}^n$ and $\{\sLi{\lambda, \gamma}\}_{i = 1}^n$ to identify data-dependent tuning parameter pairs $(\lambda,\gamma)$ that satisfies the expected risk control guarantee
\begin{equation}
\label{eq:type-1}
     \Exs \fR(\lambda) \leq \levelone \quad \text{and} \quad  \Exs \sR(\lambda, \gamma) \leq \leveltwo. 
\end{equation}
When there exists a set of feasible tuning parameter pairs, $\mathcal{R}$, that satisfy~\eqref{eq:type-1}, we are also interested in a uniform expected risk control guarantee
\begin{equation}
\label{eq:type-2}
\begin{split}
 &\Exs \sup_{(\lambda,\gamma) \in \rejset} \fR(\lambda) \leq \levelone, \quad \text{and} \\
 &\Exs \sup_{(\lambda,\gamma) \in \rejset } \sR(\lambda, \gamma) \leq \leveltwo,
 \end{split}
\end{equation}
as this would allow us to select tuning parameter pair from $\rejset$ based on certain objective function with valid expected risk control guarantee. Note that the selection of  $\lambda$ and $\gamma$ encodes the prediction set sizes. Throughout the paper, we assume $\levelone, \leveltwo$ are fixed to be in the interval $[0,1]$, which implies feasible risk control. 

With regard to two-stage risk control, one might wonder: if we can manage risk effectively in the second stage, why invest effort in controlling it in the first stage? The reason is that controlling risk in the first stage is foundational to the entire process. In the case of the ranked retrieval problem, retrieving all documents in the first stage imposes a significant computational burden on the ranking stage. Conversely, retrieving too few relevant documents in the first stage undermines the feasibility of second-stage risk control and compromises ranking quality. Thus, this paper focuses on simultaneously controlling risks at both stages. 

\subsection{Data structure in ranked retrieval problem}
Before discussing how to achieve risk control as specified in equations~\eqref{eq:type-1} and~\eqref{eq:type-2}, we first outline the structure of the data for the ranked retrieval problem. Consider a set of i.i.d calibration data points $\{ (X_i, Y_i, Z_i)\}_{i=1}^n$, with $(X_i, Y_i, Z_i) \in \mathcal{X} \times \mathcal{Y} \times \mathcal{Z}$, where $Y_i$ and $Z_i$ are the labeled responses of feature vector $X_i$ corresponding to the two stages respectively. In the ranked retrieval problem,  $X_i$ represents a query along with its associated candidate documents. We let $X_i = \{q_i, \{d_{i,j}\}_{j=1}^{N_i} \}$, where $q_i$ denotes the user query and $\{d_{i,j}\}_{j=1}^{N_i}$ denotes its associated documents with size $N_i$.
Each pair of $q_i$ and $d_{i,j}$ is associated with a ground truth relevance score $r_{i,j} \in \{0,1, \ldots,R\}$, where a higher value of $r_{i.j}$ indicates a higher relevance of $d_{i,j}$ to $q_i$. This score is only observable for the training data and is hidden for the test data. Here we denote the relevant documents with a relevance score great than $0$ in the retrieval results of $q_i$ by $Y_i = \{ d_{i,j}: r_{i,j} > 0\}$. In the ranking stage, with a focus on the ranking quality of documents with a ground truth relevance level $r_0 \in [R]$ or above,  the set of $r_0$-relevant documents for $q_i$: $\{d_{i,j}: r_{i,j} \geq r_0 \}$ is considered. Then, $Z_i$ denotes the ordered set of the $r_0$-relevant documents, sorted in descending order based on the ground truth relevance scores with ties broken arbitrarily: $Z_i = \{d_{i,(1)}, d_{i,(2)},\ldots \}$. 

Without loss of generality, we assume that each stage is associated with a model learned on the training data for all queries, denoted by  $\mretrieval$ for the retrieval model  and by $\mrank$ for the ranking model, respectively. Note that the form of the retrieval model is flexible; it can range from a simple Okapi BM25 model, which counts word occurrences, to a more complex large language model that generates embeddings for embedding-based retrieval. Typically, the model used in the retrieval stage is more efficient but less powerful than the one in the ranking stage. 
For both stages, by leveraging the pre-trained model and calibration data, we can construct prediction sets $\fConf{x;\lambda}$ for the unknown response $y \in \Yspace$ and $\sConf{x;\lambda, \gamma}$ for the unknown response $z \in \Zspace$, given a test data point $(x, y, z) \in \Xspace \times \Yspace \times \Zspace$. By employing the loss functions $\fll$ and $\sll$ for the first and second stages, respectively, we obtain:
\begin{align*}
    & \fLi{\lambda} = \fll(\fConf{X_i;\lambda}, Y_i), \quad \text{and} \\
   & \sLi{\lambda, \gamma} = \sll(\sConf{X_i; \lambda, \gamma}, Z_i).
\end{align*}
We will specify the choice of $\fConf{\cdot;\lambda}$, $\sConf{\cdot; \lambda,\gamma}$, $\fll$, and $\sll$ for ranked retrieval problem in Section~\ref{sec:app}. 

%% file: files/method.tex
\section{Two-stage risk control}
In this section, we discuss two approaches to achieve expected risk control guarantee. The first approach leverages the LTT framework \citep{aangelopoulos_ltt}, as high-probability risk control provides a stronger guarantee compared to expected risk control. The second approach extends the conformal risk control framework \citep{angelopoulos2022conformal} to accommodate the two-stage setup.

\subsection{LTT framework}

In the first stage, the value of $\lambda$ is determined by evaluating its risk function $\fR(\lambda)$ through the following hypothesis tests for each $\lambda_i \in \Lambda$:
\begin{equation}
        \Hione: \fR(\lambda_i) > \levelone  ~~\text{ vs. }~~ \Hipone: \fR(\lambda_i) \le \levelone.
\end{equation}
Given $\lambda = \lambda_i \in \Lambda$, the value of $\gamma$ is selected in the second stage by using the corresponding risk function $\sR(\lambda_i, \gamma)$. For each $\gamma_j \in \Gamma$, the following hypothesis tests are performed:
\begin{equation}
        \Hijtwo: \sR(\lambda_i, \gamma_j) > \leveltwo  ~~\text{ vs. }~~ \Hijptwo: \sR(\lambda_i, \gamma_j) \le \leveltwo.
\end{equation}
Let $\Fone=\left\{\Hione: i=1, \ldots, m\right\}$ denote the collection of all hypotheses tested in the first stage. Given $\lambda=\lambda_i \in \Lambda$, let $\Fitwo =\left\{\Hijtwo: j=1, \ldots, m\right\}$ denote the collection of all hypotheses tested in the second stage for a fixed $\lambda_i$. Finally, let $\Ftwo=\bigcup_{i=1}^m \Fitwo$ represent the complete collection of all hypotheses tested in the second stage. To conduct hypothesis testing across both stages, we aim to control the global family-wise error rate (FWER), defined as the probability of making at least one Type I error across the two families $\Fone$ and $\Ftwo$.
For each individual hypothesis $\Hione \in \Fone$ in the first stage and $\Hijtwo\in \Ftwo$ in the second stage, we use the Hoeffding-Bentkus inequality to compute $p$-values $\pione$ and $\pijtwo$, as introduced in \citep{angelopoulos2021learn}. These $p$-values are valid under their respective null hypotheses and are defined as follows:
\begin{align*}
         \pione = & \min (\exp \{-n h(\feRi \wedge \levelone, \levelone)\}, \\
        & \qquad e \mathbb{P}(\operatorname{Bin}(n, \levelone) \leq\lceil n \feRi\rceil)), \\
         \pijtwo =  & \min (\exp \{-n h(\seRij \wedge \leveltwo, \leveltwo)\}, \\
         & \qquad e \mathbb{P}(\operatorname{Bin}(n, \leveltwo) \leq\lceil n \seRij \rceil)), 
\end{align*}
where the function $h(a, b)$ is given by:
    $$
        h(a, b)=a \log \left(\frac{a}{b}\right)+(1-a) \log \left(\frac{1-a}{1-b}\right).
    $$
\begin{figure}[t]
    \centering
    \includegraphics[width=0.6\linewidth]{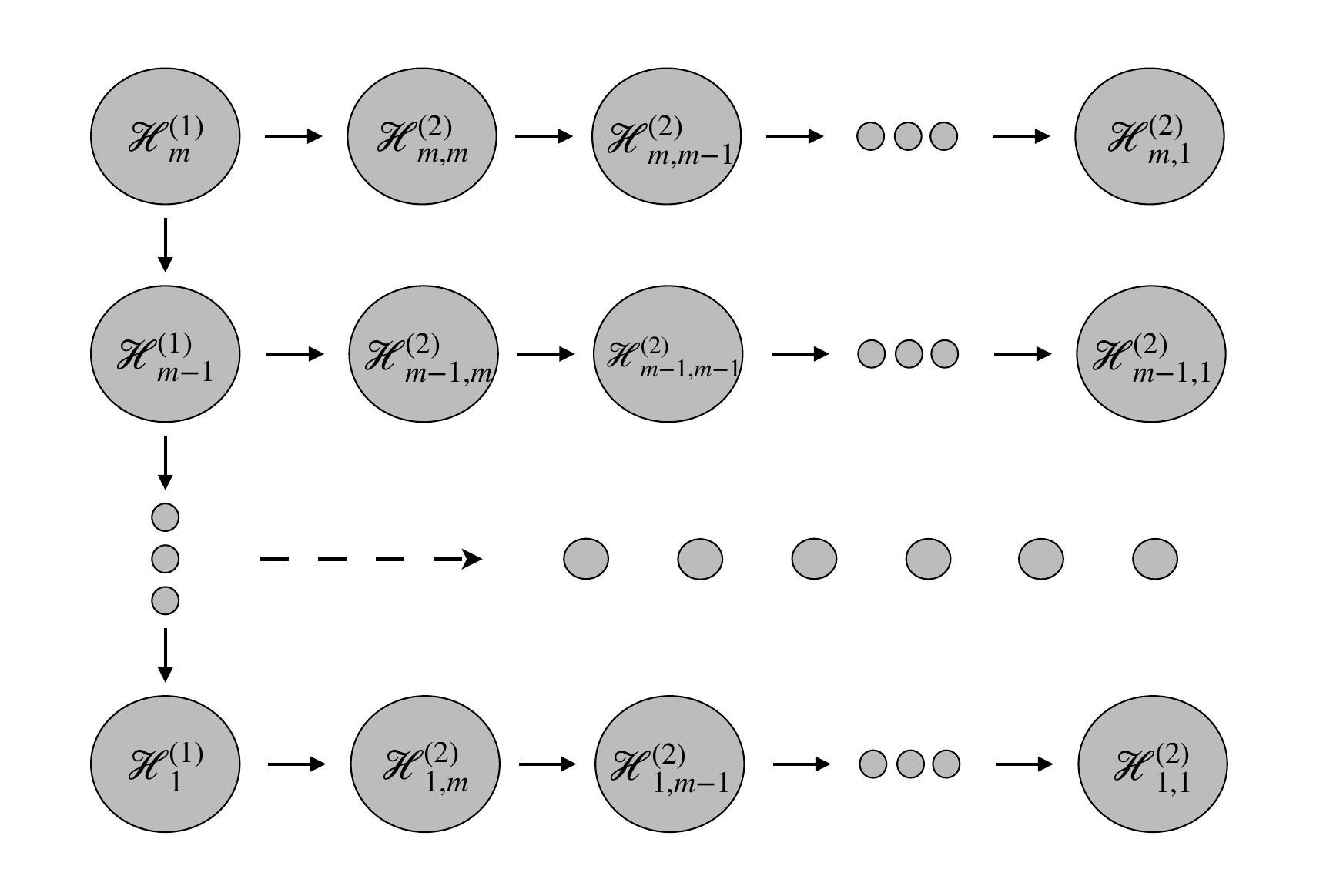}
    \caption{Graphical ordering of hypothesis tests}
    \label{fig:graph}
\end{figure}

To sequentially test the families $\Fone$ and $\Fitwo$ for each $i \in[1, m]$, one can employ the FWER controlling algorithm to obtain the tuning parameter pairs~\citep{angelopoulos2021learn}. Given that the loss function in the first stage, $\fLi{\lambda}$, is non-increasing in $\lambda$, and the loss function in the second stage, $\sLi{\lambda, \gamma}$, is non-increasing in both $\lambda$ and $\gamma$, the fixed-sequence procedure tests the hypotheses in these families in reverse order. Specifically, for $\Fone$, the $m$ hypotheses are tested sequentially in the order $\Hone_m, \Hone_{m-1}, \ldots,  \Hone_{1}$. Similarly, for $\Fitwo$, the hypotheses are tested sequentially in the order $\Htwo_{i,m}, \Htwo_{i,m-1}, \ldots, \Htwo_{i,1}$. Building on these considerations, we propose the following procedure for simultaneously testing $\Fone \cup \Ftwo$, ensuring control of the global FWER at the pre-specified level $\delta$:
\paragraph{Procedure:} 
    \begin{enumerate}
        \item \textbf{Test $\Fone$ using the Bonferroni procedure:} \\
        Apply the Bonferroni procedure to the $p$-values $\pione$ to simultaneously test $\Fone$ at level $\delta$. Let $\Rejone$ denote the index set of the rejected hypotheses:
        $$
        \Rejone=\left\{i \in[m]: \pione \leq \delta / m\right\}.
        $$
        \item \textbf{Test $\Fitwo$ using the fixed-sequence procedure:} \\
        For each $i \in \Rejone$, use the fixed-sequence procedure on the $p$-values $\pijtwo$ to simultaneously test $\Fitwo$ at level $\delta / m$. Let $\Rejtwo_i$ denote the index set of the rejected hypotheses:
        $$
        \Rejtwo_i=\left\{j \in[m]: \ptwo_{i, j^\prime} \leq \delta / m \quad \forall j^{\prime} \in[j, m]\right\}.
        $$
        \item \textbf{Determine the final set of rejected hypothesis pairs:} \\
        The set $\mathcal{R}$ of tuning parameter pairs that correspond to the rejected pairs of hypotheses $\left(\Hione, \Hijtwo \right)$ is:
        $$
        \mathcal{R}=\left\{(\lambda_i, \gamma_j): i \in \Rejone, j \in \Rejtwo_i\right\}.
        $$
    \end{enumerate}
The procedure described above can be viewed as a special case of the sequential graphical approach for multiple testing \citep{bretz2009graphical, angelopoulos2021learn}. Consequently, it strongly controls the global FWER at the pre-specified level $\delta$. We note that some other alternative global FWER-controlling procedures can be applied; a few examples are provided in the appendix. However, determining optimal procedure for a specific setting remains an open problem. In the following, we present the expected risk control guarantee for general FWER-controlling procedure.
\begin{theos}
\label{theo:ltt}
Let $\rejset$ denote the collection of tuning parameter pairs returned from a FWER controlling algorithm testing $\Fone \cup \Ftwo$ at level $\delta$. Then, we have 
\begin{equation}
    \Prob \lp \forall (\lambda,\gamma) \in \rejset: \fR(\lambda) \leq \levelone, \sR(\lambda, \gamma) \leq \leveltwo  \rp \geq 1 - \delta.
\end{equation}
Therefore, we have
\begin{equation}
\label{eq:type-2-ltt}
\begin{split}
 & \Exs \sup_{(\lambda,\gamma) \in \rejset} \fR(\lambda) \leq \levelone + \delta, \quad \text{and}  \\
 & \Exs \sup_{(\lambda,\gamma) \in \rejset } \sR(\lambda, \gamma) \leq \leveltwo + \delta.
 \end{split}
\end{equation}
\end{theos} 
\vspace{5pt}
\noindent A few comments are in order regarding the comparison with multiple risks in~\cite{angelopoulos2021learn}. The authors therein discuss a similar setting involving multiple risk functions. Specifically, consider the case where the sets of tuning parameters have size $m$, i.e., $|\Lambda| = |\Gamma| = m$. Their procedure uses the set of all parameter pairs $\{(\lambda, \gamma): \lambda \in \Lambda, \gamma \in \Gamma\}$, as input for running FWER controlling algorithm. In contrast, our approach leverages the sequential nature of the problem and the monotonicity of the risk functions, which reduces the computational burden and enhances overall effectiveness.

\subsection{Expected risk control}
\label{sec:risk-mono}
In this section, we describe methods that leverage conformal risk control framework. To proceed, we consider $\levelone, \leveltwo \in (\frac{1}{n+1}, 1]$ and introduce a few additional notations.  For the purpose of risk control in the first stage, define:
\be
\label{eq:def-hlzone}
\hlzone =\inf \left \{\lambda \in \Lambda:  \sum_{i=1}^n \fLi{\lambda} \leq (n+1)\levelone -1 \right \}.
\ee
Similarly, to ensure the feasibility of risk control in the second stage, define:
\be
\hlztwo =\inf \left \{\lambda \in \Lambda:\sum_{i=1}^n \sLi{\lambda, 1}   \leq (n+1)\leveltwo - 1 \right\}.
\ee
Lastly, for any fixed $\lambda \in \Lambda$, we define:
\be
\label{eq:def-hg}
\hgztwo(\lambda) = \inf \left \{\gamma \in \Gamma:  \sum_{i=1}^n \sLi{\lambda, \gamma} \leq (n+1)\leveltwo - 1 \right\}.
\ee
When the set in equation~\eqref{eq:def-hg} is empty, we set $\hgztwo(\lambda) = 1$.
To ensure risk control in both stages, we set the value of $\lambda$ as a linear combination of $\hlzone \vee \hlztwo$ and 1:
\be
\hlonet:= \left\lceil t (\hlzone \vee \hlztwo )+(1-t) \right\rceil_{\Lambda}.
\ee
Here $\hlonet$ is defined to be the ceiling of  $t(\hlzone \vee \hlztwo)+(1-t)$, restricted to values in $\Lambda$, and $t\in[0,1]$ is a tuning parameter. 
This formulation ensures that $\hlonet \geq \hlzone \vee \hlztwo$ for any $t$.  Using $\hlonet$, we determine the value of $\gamma$ by defining
\be
\hgtwot:= \hgztwo\left(\hlonet\right).
\ee

With the above definitions, we are ready to state our result for expected risk control.

\begin{theos}
\label{theo:two-stage-asym}
    For any $t \in$ $[0,1]$, tuning parameter pair $( \hlonet, \hgtwot )$ achieves first-stage risk control at level $\levelone$ with  a finite-sample guarantee and achieves second-stage risk control  at level $\leveltwo$ asymptotically, i.e., 
    \ben
    \begin{split}
& \Exs \fR(\hlonet) \leq \levelone, \quad \text{and}\\
& \limsup_{n \rightarrow \infty} \Exs \sR(\hlonet, \hgtwot) \leq \leveltwo.
\end{split}
\een
\end{theos}

\noindent Since $\Lambda$ and $\Gamma$ are finite sets, we further have the following corollary.
\begin{cors}
    \label{theo:cor-asym}
    Uniform risk control for both stages can be achieved in the set
    \ben
    \rejset = \lb (\lambda, \gamma) \in \Lambda \times \Gamma: \lambda = \hlonet, \gamma \geq \hgztwo(\lambda), t\in [0,1]\rb,
    \een
i.e., 
\ben
\begin{split}
& \Exs \sup_{(\lambda, \gamma) \in \rejset }\fR(\lambda) \leq \levelone, \quad \text{and} \\
&  \limsup_{n \rightarrow \infty} \Exs \sup_{(\lambda, \gamma) \in \rejset} \sR(\lambda, \gamma) \leq \leveltwo.
\end{split}
\een
\end{cors}

\subsubsection{Finite-sample second-stage risk control}
To ensure finite-sample guarantee for the second stage, a data-splitting approach can be employed. Let calibration data be divided into two non-overlapping parts with index sets $\dataone$ and $\datatwo$. Let $\levelone,\leveltwo \in ( 1/(1+|\dataone|),1]$. Define
\begin{align*}
    \tlzone =\inf \left\{\lambda \in \Lambda:  \sum_{i\in\dataone} \fLi{\lambda} \leq (|\dataone|+1)\levelone - 1\right\}.
\end{align*}
To proceed, we need to impose the following additional assumption to ensure finite-sample risk control in the second stage:
\begin{ass}
\label{ass:fea}
    There exists a known constant $\lambda_0 \leq 1$ such that $\sLi{\lambda_0, 1} \leq \leveltwo$ for $i\in[n+1]$.
\end{ass}
\noindent The above Assumption~\ref{ass:fea} enables feasibility of finite-sample second-stage risk control. Next, for $\lambda \in \Lambda\cap [\lambda_0,1]$ define
\ben
 \tgztwo(\lambda) = \inf \left\{\gamma \in \Gamma: \sum_{i\in\datatwo} \sLi{\lambda, \gamma} \leq (|\datatwo|+1)\leveltwo - 1 \right\}.
 \een
\noindent We define $\tgztwo(\lambda) = 1$ when the set is empty. For $t\in[0,1]$, we then define:
\be
\begin{split}
& \tlonet:= \left\lceil t (\tlzone \vee \lambda_0 )+(1-t) \right\rceil_{\Lambda} ,\quad \text{and} \\
& \tgtwot = \tgztwo(\tlonet).
\end{split}
\ee

\begin{theos}
\label{theo:two-stage-fin}
    For any $t \in$ $[0,1]$, tuning parameter pair $( \tlonet, \tgtwot )$ achieves finite-sample first-stage and second-stage risk control at level $\levelone$, $\leveltwo$, respectively, i.e., 
    \ben
\Exs \fR(\hlonet) \leq \levelone \quad \text{and} \quad  \Exs \sR(\hlonet, \hgtwot) \leq \leveltwo.
\een
\end{theos}
\vspace{10pt}
\noindent Note that for any $t\in[0,1]$, we have $\Exs \sR(\hlonet, \hgtwot) \leq \leveltwo$. However, there is no guarantee that $$\Exs\sup_{t\in[0,1]} \sR(\hlonet, \hgtwot) \leq \leveltwo,$$ as no monotonic relationship exists. Therefore, to obtain a finite-sample uniform risk control in the second stage, we define
\ben
\label{eq:gamma-uniform}
\bgtwo = \inf \left\{\gamma \in \Gamma:  \sum_{i\in\datatwo}   \sLi{\tlone(1), \gamma} \leq (|\datatwo|+1)\leveltwo - 1\right\}.
\een

\begin{cors}
\label{theo:cor-fin}
 Uniform risk control for both stages can be achieved in the set
 \ben
    \rejset = \lb (\lambda, \gamma) \in \Lambda \times \Gamma: \lambda = \tlonet, \gamma \geq \bgtwo, t\in [0,1]\rb,
    \een
i.e.,
    \ben
\Exs \sup_{(\lambda, \gamma) \in \rejset }\fR(\lambda) \leq \levelone \quad \text{and} \quad \Exs \sup_{(\lambda, \gamma) \in \rejset} \sR(\lambda, \gamma) \leq \leveltwo.
\een
 
\end{cors}
\vspace{5pt}
\noindent We remark that our data-splitting approach, which ensures a uniform finite-sample risk control guarantee, can be regarded as a special case of the method proposed in Section 4.3 of~\cite{angelopoulos2022conformal}.

%% file: files/application.tex
\section{Application to ranked retrieval}
\label{sec:app}
In this section, we apply our proposed methods to ranked retrieval problem. We begin by introducing the loss functions defined for each stage.
\subsection{Loss function for retrieval stage}
As defined in the Section~\ref{sec:setup}, $Y_i$ is the set of relevant documents with respect to the query $q_i$, i.e., the set of documents with ground truth relevance scores greater than $0$. To represent the set of documents fetched by $\mretrieval$ used in the retrieval stage, we define the retrieved document set $\fConf{X_i; \lambda}$ for the retrieval stage as: 
\begin{equation}
\label{eq:def-c1}
\fConf{X_i; \lambda} =\{d_{i,j}: \mretrieval(q_i, d_{i,j }) \geq 1- \lambda\},
\end{equation}
where $\lambda \in [0,1]$ and  $\mretrieval(q_i, d_{i,j })$ denotes the model score for the query-document pair $(q_i,d_{i,j})$, as computed by the retrieval model $\mretrieval$. Note that a good retrieved document set $\fConf{X_i; \lambda}$ associated with the query $q_i$ should aim to cover as many relevant documents in $Y_i$ as possible. To measure the miscoverage of  $Y_i$ by  $\fConf{X_i; \lambda}$, we define retrieval loss as: 
\begin{equation}
\label{eq:retri-loss}
\fLi{\lambda} = 1-\frac{|Y_i\cap \fConf{X_i; \lambda} |}{|Y_i|}.
\end{equation}
Note that the loss function defined in equation~\eqref{eq:retri-loss} is non-increasing, right-continuous in $\lambda$, and bounded within $[0, 1]$. This form of loss function quantifies the missed fraction of relevant documents retrieved by model $\mretrieval$. By choosing an appropriate $\lambda$, we aim to control the risk at level $\levelone$.

\subsection{Loss function for ranking stage}
Similarly, in the ranking stage, we first define the prediction set $\sConf{X_i;\lambda,\gamma}$ as:
\begin{equation}
\sConf{X_i;\lambda,\gamma} =\{d_{i,j}:  \mrank(q_i,d_{i,j}) \ge 1- \gamma   \} \cap \fConf{X_i;\lambda}.
\end{equation}
In contexts like information retrieval, search engines, and recommendation systems, nDCG \citep{Järvelin2000} is widely used to evaluate the ranking quality of algorithms or systems.  Motivated by the intuition from nDCG that a ranked list can be evaluated by rewarding relevance while considering ranking positions, we slightly modify the definition of nDCG to suit our setting, and define:\\
\begin{equation*}
\text{DCG}_{\mathrm{mod}}\left(\sConf{X_i;\lambda,\gamma}, Z_i \right) = \sum_{j = 1}^{|Z_i|}  \frac{\mathbf{1}_{\{ d_{i,{(j)}} \in \sConf{X_i;\lambda,\gamma}\}}}{\log(j+1)}.
\end{equation*}
Notably, when $\sConf{X_i;\lambda,\gamma}$ contains all the $r_0$-relevant documents, $\text{DCG}_{\mathrm{mod}}$ attains its maximum value. To normalize the $\text{DCG}_{\mathrm{mod}}$, we define the modified Ideal Discounted Cumulative Gain ($\text{iDCG}_{\mathrm{mod}}$) for query $q_i$ as:
\begin{equation*}
\text{iDCG}_{\mathrm{mod}}\left(Z_i \right) = \sum_{j=1}^{|Z_i|} \frac{1}{\log(j+1)}.
\end{equation*} 
Correspondingly, we define the modified Normalized Discounted Cumulative Gain ($\text{nDCG}_{\mathrm{mod}}$) as
\begin{equation*}
\text{nDCG}_{\mathrm{mod}}\left(\sConf{X_i;\lambda,\gamma}, Z_i \right) = \frac{\text{DCG}_{\mathrm{mod}}\left(\sConf{X_i;\lambda,\gamma} , Z_i\right)}{\text{iDCG}_{\mathrm{mod}}(Z_i)} .
\end{equation*}
The loss function for the ranking stage is then defined as
\begin{equation}
\label{eq:rank-loss}
    \sLi{\lambda, \gamma} =  1- \text{nDCG}_{\mathrm{mod}}(\sConf{X_i;\lambda,\gamma}, Z_i).
\end{equation}
The proposed loss function addresses the ranking order within the ground truth set $Z_i$. When selecting an appropriate $\gamma$ that satisfies risk control guarantee, greater weight is assigned to documents in $Z_i$ with higher ranking positions, ensuring that the most relevant documents are prioritized for inclusion in the prediction set. Note that, given $\lambda$ specified in the first stage, the ranking loss function is non-increasing, right-continuous in $\gamma$, and bounded within $[0,1]$. Our goal is to determine $\gamma$ to control the ranking loss at specified level $\leveltwo$.

\subsection{Parameter pair selection via empirical set sizes minimization}
Given a collection of tuning parameter pairs that achieve risk control at both stages, we determine the tuning parameter pair $(\hat \lambda, \hat \gamma)$ through optimization with a objective function $\mathcal{L}$:
\begin{equation}
\label{eq:optim}
(\hat \lambda, \hat \gamma) = \argmin_{(\lambda, \gamma) \in \rejset}  \mathcal{L}\lp \lambda,\gamma; \{X_i\}_{i = 1}^n \rp.
\end{equation}
In many applications (recommendation in mobile devices etc.), it is desirable to produce a smaller prediction set in the second stage. Therefore, in this paper, we consider objective function
\begin{align*}
    \mathcal{L}\lp \lambda,\gamma; \{X_i\}_{i = 1}^n \rp =  \frac{1}{n}\sum_{i = 1}^n |\sConf{X_i;\lambda,\gamma} \}|.
\end{align*}
We comment that, the choice of the loss function should be driven by the specific problem at hand. For instance, in scenarios where the objective is to reduce computational cost during the ranking stage, a smaller prediction set size in the retrieval stage reduces the number of documents evaluated by the computationally intensive ranking model.  To address this, the term $\frac{1}{n}\sum_{i = 1}^n |\fConf{X_i;\lambda}|$ can be incorporated into the loss function $\mathcal{L}$, thereby systematically mitigating the computational burden.

\subsection{Experiments}
We address the ranked retrieval problem using three approaches: the learn-then-test framework (LTT), two-stage conformal risk control (tCRC), and two-stage conformal risk control with data splitting (tCRC-s\footnote{While the required information $\lambda_0$ for tCRC-s is unknown in practice, we use the calibration data with index set $\dataone$ to obtain an estimate. }). The tuning parameter pairs $(\hat \lambda, \hat \gamma)$ for these methods are determined by equation~\eqref{eq:optim}. Accordingly, these selected tuning parameter pairs ensure expected risk control guarantees, as established in Theorem~\ref{theo:ltt}, Corollary~\ref{theo:cor-asym}, and Corollary~\ref{theo:cor-fin}, respectively. Our experiments are conducted on two datasets: MSLR-Web, and Yahoo LTRC. For each dataset and experiment, we split the data into a calibration set and a test set, and then apply different methods to the data. The calibration set is used to determine the tuning parameter pair, while the test set is used for evaluation. Using the test data and the tuning parameter pair computed from the calibration data, we compute the following quantities for evaluation: empirical risks in two stages (Risk $(1)$ represents empirical risk in the first stage, Risk $(2)$ represents empirical risk in the second stage), average prediction set size in the second stage, average recall for documents with a relevance level greater than $2$, average recall for documents with a relevance level equal to $1$, and average precision for documents with relevance level greater than $1$. Then, we replicate each experiment $10$ times and report the average results.

\begin{table*}[!h]\centering
\small
\caption{Summary table for dataset MSLR}\label{tab:mslr}
\begin{tabular}{|c|c|c|c|c|c|c|c|c|}\toprule
$(\levelone, \leveltwo)$ & Method & Risk (1) & Risk (2) & Set size &  Recall ($\geq 2$) & Recall (1) & Precision   \\\cmidrule(lr){1-8}
 & tCRC  &  0.0069 & 0.0982 & 67.86 & 0.9537 & 0.8550 & 0.7504 \\
$(0.1,0.1)$& tCRC-s  & 0.0986 & 0.0969 & 77.41& 0.9279& 0.8790& 0.6673 \\
& LTT & 0.0003 & 0.0772 & 70.25 & 0.9642 & 0.8857 & 0.7463 \\\cmidrule(lr){1-8}
 & tCRC &  0.0058& 0.0994 & 68.08 & 0.9532 & 0.8533  & 0.7511 \\
$(0.01,0.1)$& tCRC-s & 0.0078 & 0.0967 & 68.44 & 0.9537 & 0.8577  &  0.7503\\
& LTT & 0.0002 & 0.0765 & 70.66 & 0.9646& 0.8864& 0.7467 \\\cmidrule(lr){1-8}
 & tCRC & 0.0051 & 0.2006 & 57.50 & 0.9084 & 0.7017 & 0.7578\\
$(0.1,0.2)$ & tCRC-s   &  0.0972& 0.1971& 58.28 & 0.8825 & 0.7300& 0.7575 \\
& LTT & 0.0046 & 0.1718 & 60.22 & 0.9221 & 0.7440 & 0.7578 \\
\bottomrule
\end{tabular}
\end{table*}

\begin{figure}[!b]
    \centering
    \includegraphics[width=\linewidth]{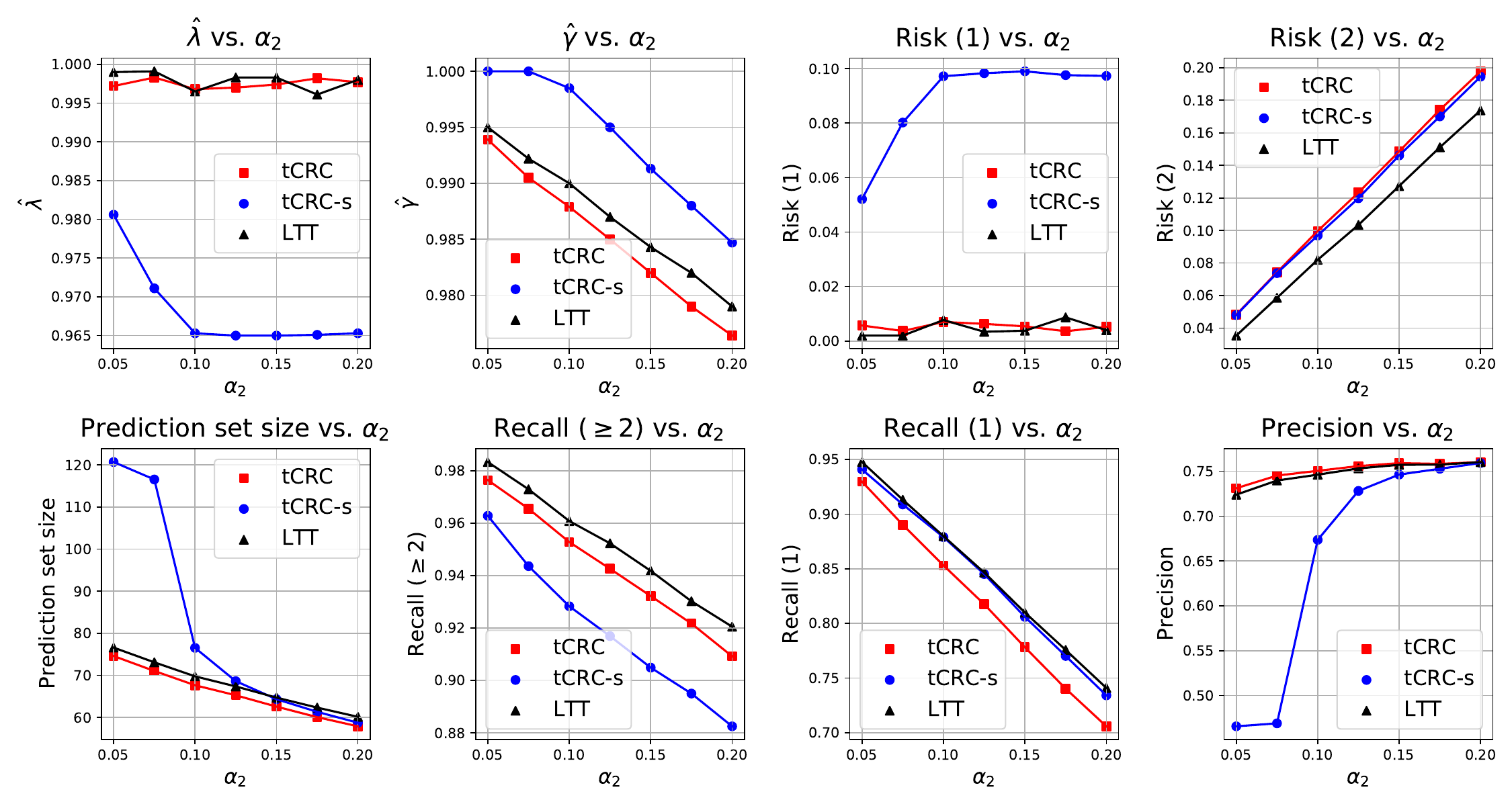}
    \caption{MSLR dataset}
    \label{fig:mslr}
\end{figure}

In Table~\ref{tab:mslr}, we summarize results for the dataset MSLR. From the table, we observe that both Recall $(\geq 2)$ and Recall $(1)$ exhibit relatively high values, indicating that the prediction sets effectively cover a substantial proportion of relevant documents. Additionally, Recall $(\geq 2)$ consistently achieves higher values than Recall $(1)$. This confirms that our proposed ranking loss effectively prioritizes documents with higher relevance levels, resulting in prediction sets that are more likely to include highly relevant documents. In the table, we vary the risk levels and consider $(\levelone, \leveltwo) \in \{ (0.1, 0.1), (0.01, 0.1), (0.1, 0.2) \}$. Notably, LTT exhibits the lowest Risk $(1)$ and Risk $(2)$ values, indicating that LTT is the most conservative approach. This can also be explained by the fact that the size of the feasible set $\rejset$ for LTT is, on average, the smallest. Conversely, we observe that tCRC achieves the smallest prediction set size after tuning parameter selection via equation~\eqref{eq:optim}, attributed to its larger feasible set $\rejset$. Notably, compared with tCRC-s, tCRC maintains a larger set size in the second stage but achieves a smaller Recall $(\geq2)$. While this seems counterintuitive, it can be explained by that fact that tCRC-s has a larger risk in the first stage. This is achieved by using a smaller $\hat{\lambda}$, which results in fewer documents with higher relevance levels being retrieved in the first stage, ultimately impacting the performance in the second stage. When $\alpha_1$ is reduced from $0.1$ to $0.01$, the metrics of tCRC and LTT exhibit limited variation, whereas tCRC-s demonstrates greater sensitivity to this change. This increased sensitivity can be attributed to the construction of the feasible set by tCRC-s. Additionally, when $\alpha_2$ is increased from $0.1$ to $0.2$, we observe that Risk $(2)$ slightly exceeds the target risk level of $0.2$. This deviation can be explained by approximation error and the asymptotic validity of tCRC, as established in Corollary~\ref{theo:cor-asym}.

\begin{figure}[!h]
    \centering
    \includegraphics[width=\linewidth]{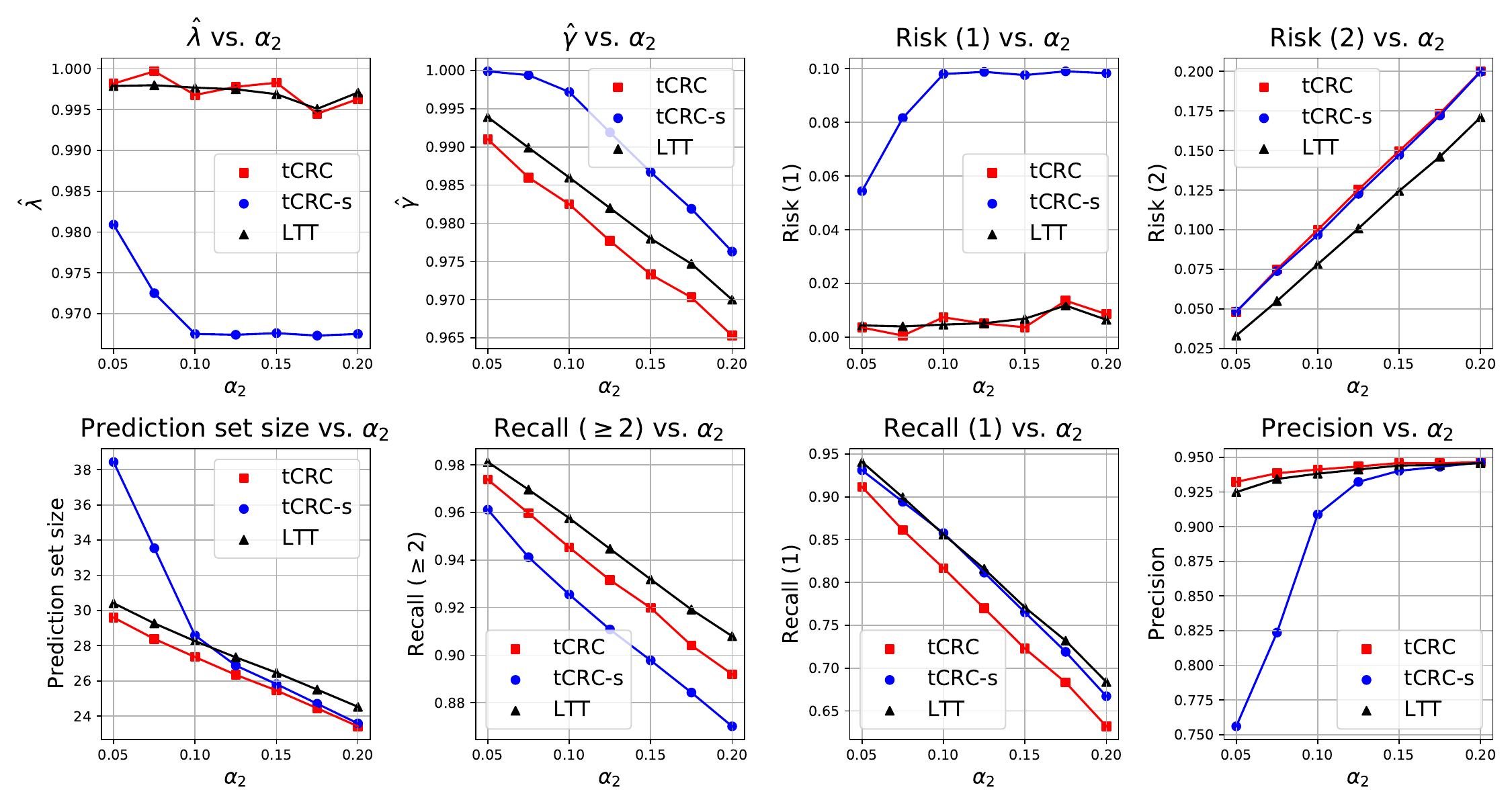}
    \caption{Yahoo dataset}
    \label{fig:yahoo}
\end{figure}

In Figure~\ref{fig:mslr}, we fix $\alpha_1 = 0.1$ and vary $\alpha_2$ within the set $\{0.05, 0.075, 0.1, 0.125, 0.15, 0.175, 0.2\}$. All three methods effectively manage risk within the desired levels. Notably, as $\alpha_2$ varies, tCRC achieves the smallest prediction set size in the second stage, LTT attains the highest recall rate, and both tCRC and LTT demonstrate superior precision. In practice, we recommend using LTT or tCRC. When the sample size is small, tCRC-s may suffer from lower sample efficiency, compounded by the challenge of the unknown parameter $\lambda_0$. Lastly, we present similar results for the Yahoo dataset in Table~\ref{tab:yahoo}, and Figure~\ref{fig:yahoo}. 

\begin{table*}[!tp]\centering
\small
\caption{Summary table for dataset Yahoo}\label{tab:yahoo}
\begin{tabular}{|c|c|c|c|c|c|c|c|c|}\toprule
$(\levelone, \leveltwo)$ & Method & Risk (1) & Risk (2) & Set size &  Recall ($\geq 2$) & Recall (1) & Precision   \\\cmidrule(lr){1-8}
 & tCRC  &  0.0053 & 0.1001 & 27.50 & 0.9458 & 0.8157 & 0.9414 \\
$(0.1,0.1)$ & tCRC-s  & 0.0985 & 0.0974 & 28.80 & 0.9246 & 0.8577 & 0.9069
 \\
& LTT & 0.0052 & 0.0787 & 28.40 & 0.9572 & 0.8556 & 0.9383
 \\\cmidrule(lr){1-8}
 & tCRC &  0.0025 & 0.0984 & 27.46 & 0.9466 & 0.8187 & 0.9414
 \\
$(0.01,0.1)$& tCRC-s & 0.0078 & 0.0966 & 27.56 & 0.9464 & 0.8245 & 0.9405
\\
& LTT & 0.0019 & 0.0748 & 28.43 & 0.9599 & 0.8610 & 0.9381
 \\\cmidrule(lr){1-8}
 & tCRC & 0.0144 & 0.2014 & 23.36 & 0.8885 & 0.6325 & 0.9480
\\
$(0.1,0.2)$ & tCRC-s   &  0.0966 & 0.1991 & 23.60 & 0.8698 & 0.6682 & 0.9477
 \\
& LTT & 0.0047 & 0.1722 & 24.49 & 0.9070 & 0.6823 & 0.9472
\\
\bottomrule
\end{tabular}
\end{table*}

%% file: files/conclusion.tex
\section{Conclusion \& Discussion}
In this paper, we study expected risk control under a two-stage setup. We propose methods aimed at simultaneously controlling risk, and establish theoretical guarantees for these methods. The effectiveness of our approach is validated through experiments on two large-scale, widely used datasets: MSLR-Web and Yahoo LTRC. Below, we discuss several extensions of the current framework.
\paragraph{Non-monotone risk function}
When risk function is non-monotone, the following monotonization procedure can be applied. At the first stage, the loss function is modified as follows:
\ben
\tfLi{\lambda} = \sup_{\lambda^\prime \in [\lambda,1]\cap \Lambda }\fLi{\lambda^\prime}.
\een
At the second stage, we define the modified loss function as
\ben
\tsLi{\lambda, \gamma} = \sup_{ \small\substack{\lambda^\prime \in [\lambda,1]\cap \Lambda, \\ \gamma^\prime \in  [\gamma,1]\cap \Gamma}} \sLi{\lambda^\prime, \gamma^\prime}.
\een
With these modifications to the loss functions, the results from the previous sections can be directly applied.

\paragraph{Multiple-stages}
We extend the problem to the scenario where there are $K > 2$ stages. Let $\theta \in \R^{K}$ denote the parameter of interest. At the $k$-th stage, the loss function for sample $i$ is denoted as $L^{\scalebox{0.6}{(k)}}_i(\theta_{1:k})$, where $\theta_{1:k}$ is shorthand for $(\theta_1,\ldots, \theta_k)$. The sequential nature of the problem determines that loss function at stage $k$ only involves parameters $\theta_{1:k}$. The goal is to determine parameter set $\theta$ that satisfies
\ben
\Prob \lp \forall k \in [K], L^{\scalebox{0.6}{(k)}}_i(\theta_{1:k}) \leq \levelk, \rp \geq 1 - \delta,
\een
or alternatively,
\ben
\Exs L^{\scalebox{0.6}{(k)}}_i(\theta_{1:k}) \leq  \levelk \quad 
\text{for all $k \in [K]$}.
\een
When loss functions at different stages are monotonic in each parameter $\theta_i$ with $i \in [K]$, one can develop analogous approaches to achieve risk control with high probability and expected risk control.

%% file: files/appendix.tex
\newpage

\appendix

\begin{center}
    \Large Two-stage Risk Control with Application to Ranked Retrieval Technical Appendix
\end{center}

\vskip 5pt

\section{Theoretic Result}
\subsection{Proof of Theorem 2}
In Section 3.2, we define $\hlzone$, $\hlztwo$ and $\hgztwo$. Here, we  introduce the corresponding population-level quantities. We first define
\ben
\begin{split}
& \lzone=\inf \left\{\lambda \in \Lambda: \sR(\lambda) \leq \levelone\right\}, \quad \text{and} \\
& \lztwo =\inf \left\{\lambda \in \Lambda: \sR(\lambda, 1) \leq \leveltwo\right\}.
\end{split}
\een
For any fixed $\lambda \in \Lambda$, define
\ben
\gztwo(\lambda) = \inf \left\{\gamma \in \Gamma: \sR(\lambda, \gamma) \leq \leveltwo\right\}.
\een
We set $\gztwo(\lambda) = 1$ when the set is empty. For any $t \in [0,1]$, define
\ben
\lonet := \lceil t(\lzone \vee \lztwo ) + (1-t) \rceil_{\Lambda}.
\een


\paragraph{Proof.}
By the strong law of large numbers, for any $\lambda \in \Lambda$, we have:
\ben
\frac{1}{n+1} \sum_{i=1}^n \fLi{\lambda}+\frac{1}{n+1} \xrightarrow{\text { a.s. }} \fR(\lambda).
\een
Since $\Lambda \subset[0,1]$ is finite, it follows that
\be
\label{eq:con-1}
 \lim_{n\to \infty}  \indic_{\{\hlzone \geq \lzone\}}  = 1 \quad \text{a.s.} 
\ee
Similarly, we obtain
\be
\label{eq:con-2}
\lim_{n\to \infty}  \indic_{ \{\hlztwo \geq \lztwo\}}  = 1 \quad \text{a.s.} 
\ee
And, for any $\lambda \in \Lambda \cap [\lztwo,1]$, we have:
\be
\label{eq:con-3}
\lim_{n\to \infty}  \indic_{ \{\hgztwo(\lambda) \geq \gztwo(\lambda)\}}  = 1 \quad \text{a.s.} 
\ee
Moreover, since $\Lambda \cap [\lztwo,1]$ is a finite set, we have
\be
\label{eq:con-33}
\lim_{n\to \infty} \indic_{ \left\{ \inf_{\lambda \in \Lambda \cap [\lztwo,1]} \left( \hgztwo(\lambda) - \gztwo(\lambda)  \right)  \geq 0 \right\}} = 1\quad \text{a.s.}
\ee

For the first stage, risk control is straightforward: since $\hlonet \geq \hlzone$ for all $t \in[0,1]$, 
$\Lambda$ is finite, and $\fLn{\lambda}$ is non-increasing in $\lambda$, we can apply Theorem 1 from \citep{angelopoulos2022conformal} to obtain
\ben
\Exs \fR(\hlonet) = \Exs \fLn{\hlonet} \leq \Exs \fLn{\hlzone} \leq \levelone.
\een

For the second stage risk control,  following equations~\eqref{eq:con-1},~\eqref{eq:con-2}, and~\eqref{eq:con-33}, we have that when $n$ is sufficiently large,
\ben 
\sR\left(\hlonet, \hgtwot \right) \stackrel{(i)}{\leq} \sR\left(\hlonet, \gztwo(\hlonet)\right) \stackrel{(ii)}{\leq}   \leveltwo.
\een
Inequality $(i)$ follows from the fact that, with probability one, when $n$ is sufficiently large, $\hlonet\geq \lzone \vee \lztwo$, and $\hgztwo(\lambda)\geq \gztwo(\lambda)$. Moreover, the validity of inequality $(ii)$ can be observed from the definition of $\gztwo$. Subsequently, we have 
\ben
\limsup_{n\to \infty} \sR\left(\hlonet, \hgtwot \right) \leq \leveltwo,
\een
which leads to 
\be
\begin{split}
& \limsup_{n\to \infty} \Exs \sR\left(\hlonet, \hgtwot \right) \\
&\leq \Exs  \limsup_{n\to \infty} \sR\left(\hlonet, \hgtwot \right) \leq \leveltwo.
\end{split}
\ee
The first inequality displayed above is a direct application of Fatou's lemma. With this, we complete the proof of Theorem 2.

\subsection{Proof of Corollary 1}
Following the proof of Theorem 2, we directly have
\ben
 \Exs \sup_{t\in[0,1]} \fR(\hlonet) \leq  \Exs \fR(\hlone(1)) \leq \levelone.
\een
Similarly, as demonstrated in the proof of Theorem 2, the following holds almost surely for sufficiently large $n$:
\ben 
\begin{split}
& \sup_{(\lambda, \gamma) \in \rejset} \sR(\lambda, \gamma) \leq   \sup_{t\in [0,1]} \sR\left(\hlonet, \hgtwot \right)\\
& \leq \sup_{\lambda \in \Lambda \cap [\lztwo, 1]} \sR\left(\lambda, \gztwo(\lambda)\right) \leq   \leveltwo.
\end{split}
\een
With a similar proof as in the proof of Theorem 2, we complete the proof of Corollary 1.

\subsection{Proof of Theorem 3}
For the first-stage risk control, we apply Theorem 1 in~\citep{angelopoulos2022conformal}, and have
\begin{equation*}
    \Exs \fLn{\tlonet} \leq \Exs  \fLn{\tlzone} \leq \levelone.
\end{equation*}
For the second-stage risk control, by the tower property of conditional expectation, we have
\begin{equation*}
    \Exs \sLn{\tlonet, \tgtwot} = \Exs \left\{  \Exs \left\{ \sLn{\tlonet, \tgtwot} | \Data_1 \right\} \right\},
\end{equation*}
where $\Data_1 = \{ \fLi{\lambda}\}_{i\in\dataone} \cup \{\sLi{\lambda,\gamma}\}_{i\in\dataone}$. Here by conditioning on the $\Data_1$, $\tlonet$ can be viewed as a constant. Note that, by definition, $\tlonet \geq \lambda_0$ and $\sLi{\tlonet, 1} \leq \leveltwo$. Therefore, we can apply Theorem 1 in~\citep{angelopoulos2022conformal} to the loss functions $\{\sLi{\tlonet, \gamma}\}_{i \in \datatwo}$, and have 
\begin{equation*}
    \Exs \left\{ \sLn{\tlonet, \tgtwot} \mid \Data_1 \right\} \leq \leveltwo.
\end{equation*} 
Lastly, by taking expectation, we complete the proof of Theorem 3. 

\subsection{Proof of Corollary 2}
Note that 
\ben
 \Exs  \sup_{(\lambda,\gamma) \in \rejset} \fLn{\lambda} = \Exs  \fLn{\tlonet} \leq \levelone.
\een
For the second stage, we have
\begin{align*}
 \Exs   \sup_{(\lambda,\gamma) \in \rejset} \sR\left(\lambda, \gamma \right) &  =  \Exs  \sR(\tlone(1), \bgtwo)   \\
 & = \Exs  \sLn{\tlone(1), \bgtwo}.
\end{align*}
Conditioned on $\Data_1$, we make use of Theorem 1 in~\citep{angelopoulos2022conformal} to analyze function $ \sLi{\tlone(1), \gamma}$, and obtain
\ben
\Exs  \sLn{\tlone(1), \bgtwo} \leq \leveltwo.
\een
Lastly, by taking the expectation, we complete the proof.

\newpage

\section{Alternative procedures for controlling FWER}

    \paragraph{Procedure 1:}
    Let $w \in(0,1)$ denote a tuning parameter.
    \begin{enumerate}
        \item \textbf{Test $\Fone$ using the fixed-sequence procedure:} 
        \begin{itemize}
            \item Start with hypothesis $\Hone_m$, testing it at level $\delta$. If $\pone_m \leq \delta$, reject it and proceed to test $\Hone_{m-1}$ at level $w \delta$. Otherwise, stop testing.
            \item For $i=m-1, \ldots, 1$, test $\Hione$ at level $w^{m-i} \delta$. If $\pione \leq w^{m-i} \delta$, reject it and proceed to test $\Hone_{i-1}$ at level $w^{m-i+1} \delta$. Otherwise, stop testing. 
        \end{itemize}
        Define the index set of rejected hypotheses as:
        $$
       \Rejone=\left\{i \in [m]: \pone_{i^\prime} \leq w^{m- i^\prime}\delta \quad \forall i^{\prime} \in[i, m]\right\}.
        $$
        \item \textbf{Test $\Ftwo$ using the fixed-sequence procedure:} \\
        For each $i \in \Rejone$, use the fixed-sequence procedure on the $p$-values $\pijtwo$ to simultaneously test $\Fitwo$ at level $(1-w)w^{m-i}\delta$. Let $\Rejtwo_i$ denote the index set of the rejected hypotheses:
        $$
        \Rejtwo_i=\left\{j \in [m]: p_{i, j^{\prime}}^{(2)} \leq (1-w)w^{m-i}\delta \quad \forall j^{\prime} \in[j, m]\right\}.
        $$
        \item \textbf{Determine the final set of rejected hypothesis pairs:} \\
         The set $\mathcal{R}$ of tuning parameter pairs that correspond to the rejected pairs of hypotheses $\left(\Hione, \Hijtwo\right)$ is:
         $$\mathcal{R}=\left\{(\lambda_i, \gamma_j): i \in \Rejone, j \in \Rejtwo_i\right\}.$$
    \end{enumerate}
    
    \vskip 5pt
    
  \paragraph{Procedure 2:} 
    \begin{enumerate}
        \item \textbf{Test $\Fone$ using the Bonferroni procedure:} \\
        Apply the Bonferroni procedure to the $p$-values $\pione$ to simultaneously test $\Fone$ at level $\delta$. Let $\Rejone$ denote the index set of the rejected hypotheses:
        $$
        \Rejone=\left\{i \in[m]: \pione \leq \delta / m\right\}.
        $$
        \item \textbf{Test $\Fitwo$ using the Bonferroni procedure:} \\
        For each $i \in \Rejone$, use the Bonferroni procedure on the $p$-values $\pijtwo$ to simultaneously test $\Fitwo$ at level $\delta / m$. Let $\Rejtwo_i$ denote the index set of the rejected hypotheses:
        $$
        \Rejtwo_i=\left\{j \in[m]: \pijtwo \leq \delta / m^2 \right\}.
        $$
        \item \textbf{Determine the final set of rejected hypothesis pairs:} \\
        The set $\mathcal{R}$ of tuning parameter pairs that correspond to the rejected pairs of hypotheses $\left(\Hione, \Hijtwo\right)$ is:
        $$
         \mathcal{R}=\left\{(\lambda_i, \gamma_j): i \in \Rejone, j \in \Rejtwo_i\right\}.
        $$
    \end{enumerate}
    
  \paragraph{Procedure 3:} Let $w \in(0,1)$ denote a tuning parameter.
    \begin{enumerate}
        \item \textbf{Test $\Fone$ using the fixed-sequence procedure:}\\
        \begin{itemize}
            \item Start with hypothesis $\Hone_m$, testing it at level $\delta$. If $\pone_m \leq \delta$, reject it and proceed to test $\Hone_{m-1}$ at level $w \delta$. Otherwise, stop testing.
            \item For $i=m-1, \ldots, 1$, test $\Hione$ at level $w^{m-i} \delta$. If $\pione \leq w^{m-i} \delta$, reject it and proceed to test $\Hone_{i-1}$ at level $w^{m-i+1} \delta$. Otherwise, stop testing. 
        \end{itemize}
        Define the index set of rejected hypotheses as:
        $$
        \Rejone=\left\{i \in[m]: \pone_{i^\prime} \leq w^{m-i^\prime}\delta \quad \forall i^{\prime} \in[i, m]\right\}.
        $$
        \item \textbf{Test $\Fitwo$ using the Bonferroni procedure:} \\
        For each $i \in \Rejone$, use the Bonferroni procedure on the $p$-values $\pijtwo$ to simultaneously test $\Fitwo$ at level $(1-w)w^{m-i}\delta$. Let $\Rejtwo_i$ denote the index set of the rejected hypotheses:
        $$
        \Rejtwo_i=\left\{j \in[m]: \pijtwo \leq (1-w)w^{m-i}\delta/m \right\}.
        $$
        \item \textbf{Determine the final set of rejected hypothesis pairs:} \\
        The set $\mathcal{R}$ of tuning parameter pairs that correspond to the rejected pairs of hypotheses $\left(\Hione, \Hijtwo\right)$ is:
        $$
         \mathcal{R}=\left\{(\lambda_i, \gamma_j): i \in \Rejone, j \in \Rejtwo_i\right\}.
        $$
    \end{enumerate}

\section{Experiment details}

\subsection{Datasets}
In this study, we evaluate our proposed methods on two well-known public datasets for ranked retrieval problems: the MSLR-Web dataset\cite{} \footnote {This data can be downloaded at https://www.microsoft.com/en-us/research/project/mslr/ } and the Yahoo LTRC dataset\footnote {This data can be downloaded at https://webscope.sandbox.yahoo.com/catalog.php?datatype=c }. Both datasets consist of feature vectors extracted from query-document pairs along with relevance judgment labels. 

The MSLR-WEB dataset \citep{DBLP:journals/corr/QinL13}, a large-scale Learning-to-Rank dataset released by Microsoft Research, is curated through a commercial web search engine (Microsoft Bing). It comprises 10K queries, associated with 1.2 million documents. Each query-document pair is represented by a 136-dimensional feature vector, normalized to be in the range [0, 1]. The relevance judgments are obtained from a retired labeling set of Microsoft Bing, which takes 5 values from 0 (irrelevant) to 4 (perfectly relevant). 

The Yahoo! Learning to Rank Challenge dataset \citep{pmlr-v14-chapelle11a} consists of 172,870 documents spanning 6,330 queries, is sampled from query logs of the Yahoo! search engine. Each query-document pair is represented by a 700-dimensional feature vector, normalized to be in the range [0, 1]. Similarly, each document is given a relevance judgment with respect to the query, ranging from  from 0 (least relevant) to 4 (most relevant).

For both datasets, we train their retrieval and ranking models on respective holdout training sets, where we do not particularly tune the models for optimal ranking results, as it is not the primary focus of this work. For the retrieval model, we train a three-layer MLP, with 128 and 32 neurons in the hidden layers. For the ranking model, we utilize the LambdaRank model \citep{CBurges2006} implemented by the open-source Pytorch package PT-Rank \citep{yu2020ptranking}. This choice of models emulates real search engine practices, where the retrieval stage employs a lightweight model for efficiency, while the ranking stage adopts a more sophisticated model for quality.

\subsection{Details}
In the experiments, we consider the following choices:
\begin{itemize}
    \item $\Lambda = \Gamma = [0.950, 0.951, \ldots, 1.000]$: a list from $0.950$ to $1.000$ with step size $0.001$
    \item $\delta = 0.01$ for LTT method
\end{itemize}
\newcommand{\datatest}{\mathcal{I}_{test}}
\newcommand{\hlam}{\hat{\lambda}}
\newcommand{\hgam}{\hat{\gamma}}
We describe the calculation of evaluation metrics for selected tuning parameter pair $(\hat{\lambda}, \hat{\gamma})$, derived from calibration data. Let $\datatest$ denote the index set corresponding to the test data. The evaluation metrics are defined as follows:
\begin{itemize}
    \item \textbf{Risk $(1)$:} $$\frac{1}{|\datatwo|} \sum_{i\in \datatest} \fLi{\hlam}$$
    \item \textbf{Risk $(2)$:} $$\frac{1}{|\datatwo|} \sum_{i\in \datatest} \sLi{\hlam, \hgam}$$
    \item \textbf{Set size:} $$\frac{1}{|\datatest|} \sum_{i\in\datatest} |\sConf{X_i; \hlam,\hgam}| $$
    \item \textbf{Recall $(\geq2)$:}  $$\frac{1}{|\datatest|}  \sum_{i\in \datatest} \frac{|\sConf{X_i; \hlam,\hgam}| \cap |\{d_{i,j}: r_{i,j} \geq 2\}|}{ |\{d_{i,j}: r_{i,j} \geq 2\}| }$$
    \item \textbf{Recall $(1)$:}  $$\frac{1}{|\datatest|}  \sum_{i\in \datatest} \frac{|\sConf{X_i; \hlam,\hgam}| \cap |\{d_{i,j}: r_{i,j} = 1\}|}{ |\{d_{i,j}: r_{i,j} = 1\}| }$$
    \item \textbf{Precision:}  $$\frac{1}{|\datatest|}  \sum_{i\in \datatest} \frac{|\sConf{X_i; \hlam,\hgam}| \cap |\{d_{i,j}: r_{i,j} \geq 1\}|}{|\sConf{X_i; \hlam,\hgam}|}$$
\end{itemize}

%% file: main.bbl
\begin{thebibliography}{29}
\providecommand{\natexlab}[1]{#1}
\providecommand{\url}[1]{\texttt{#1}}
\expandafter\ifx\csname urlstyle\endcsname\relax
  \providecommand{\doi}[1]{doi: #1}\else
  \providecommand{\doi}{doi: \begingroup \urlstyle{rm}\Url}\fi

\bibitem[Angelopoulos et~al.(2023)Angelopoulos, Krauth, Bates, Wang, and Jordan]{Angelopoulos23}
A.N. Angelopoulos, K.~Krauth, S.~Bates, Y.~Wang, and M.I. Jordan.
\newblock Recommendation systems with distribution-free reliability guarantees.
\newblock In \emph{Symposium on Conformal and Probabilistic Prediction with Applications (COPA), 2023}, 2023.

\bibitem[Angelopoulos and Bates(2021)]{Angelopoulos2021}
Anastasios~N. Angelopoulos and Stephen Bates.
\newblock {A gentle introduction to conformal prediction and distribution-free uncertainty quantification}.
\newblock \emph{arXiv:2107.07511}, 2021.

\bibitem[Angelopoulos et~al.(2021{\natexlab{a}})Angelopoulos, Bates, Cand{\`e}s, Jordan, and Lei]{angelopoulos2021learn}
Anastasios~N Angelopoulos, Stephen Bates, Emmanuel~J Cand{\`e}s, Michael~I Jordan, and Lihua Lei.
\newblock Learn then test: Calibrating predictive algorithms to achieve risk control.
\newblock \emph{arXiv preprint arXiv:2110.01052}, 2021{\natexlab{a}}.

\bibitem[Angelopoulos et~al.(2021{\natexlab{b}})Angelopoulos, Bates, Candès, Jordan, and Lei]{aangelopoulos_ltt}
Anastasios~N. Angelopoulos, Stephen Bates, Emmanuel~J. Candès, Michael~I. Jordan, and Lihua Lei.
\newblock Learn then test: Calibrating predictive algorithms to achieve risk control.
\newblock \emph{arXiv preprint arXiv:2110.01052}, 2021{\natexlab{b}}.

\bibitem[Angelopoulos et~al.(2024)Angelopoulos, Bates, Fisch, Lei, and Schuster]{angelopoulos2022conformal}
Anastasios~N Angelopoulos, Stephen Bates, Adam Fisch, Lihua Lei, and Tal Schuster.
\newblock Conformal risk control.
\newblock \emph{ICLR}, 2024.

\bibitem[Baeza-Yates and Ribeiro-Neto(1999)]{rbyates_99}
R.~Baeza-Yates and B.~Ribeiro-Neto.
\newblock \emph{Modern Information Retrieval}.
\newblock ACM Press / Addison-Wesley, 1999.

\bibitem[Bretz et~al.(2009)Bretz, Maurer, Brannath, and Posch]{bretz2009graphical}
Frank Bretz, Willi Maurer, Werner Brannath, and Martin Posch.
\newblock A graphical approach to sequentially rejective multiple test procedures.
\newblock \emph{Statistics in medicine}, 28\penalty0 (4):\penalty0 586--604, 2009.

\bibitem[Burges et~al.(2005)Burges, Shaked, Renshaw, Lazier, Deeds, Hamilton, and Hullender]{cburges_icml05}
C.~J. Burges, T.~Shaked, E.~Renshaw, A.~Lazier, M.~Deeds, N.~Hamilton, and G.~Hullender.
\newblock Learning to rank using gradient descent.
\newblock In \emph{Proceedings of the 22nd international conference on Machine learning}, 2005.

\bibitem[Burges et~al.(2006)Burges, Ragno, and Le.]{CBurges2006}
Christopher~J.C. Burges, Robert Ragno, and Quoc~Viet Le.
\newblock Learning to rank with nonsmooth cost functions.
\newblock In \emph{Proceedings of NIPS conference, 2006}, 2006.

\bibitem[Cao et~al.(2007)Cao, Qin, Liu, Tsai, and Li]{zcao_msr_tr}
Zhe Cao, Tao Qin, Tie-Yan Liu, Ming-Feng Tsai, and Hang Li.
\newblock Learning to rank: From pairwise approach to listwise approach.
\newblock In \emph{MSR-TR-2007-40}, 2007.

\bibitem[Chapelle and Chang(2011)]{pmlr-v14-chapelle11a}
Olivier Chapelle and Yi~Chang.
\newblock Yahoo! learning to rank challenge overview.
\newblock In \emph{Proceedings of the Learning to Rank Challenge}, volume~14 of \emph{Proceedings of Machine Learning Research}. PMLR, 2011.

\bibitem[Chu and Ghahramani(2005)]{wchu_icml05}
W.~Chu and Z.~Ghahramani.
\newblock Preference learning with gaussian processes.
\newblock In \emph{Proceedings of the 22nd international conference on Machine learning}, 2005.

\bibitem[Crammer and Singer(2001)]{NIPS2001_5531a583}
Koby Crammer and Yoram Singer.
\newblock Pranking with ranking.
\newblock In \emph{Proceedings of NIPS conference}, 2001.

\bibitem[Freund et~al.(2003)Freund, Iyer, Schapire, and Singer]{yfreund_jmlr03}
Y.~Freund, R.~Iyer, R.~E. Schapire, and Y.~Singer.
\newblock An efficient boosting algorithm for combining preferences.
\newblock In \emph{Journal of Machine Learning Research}, 2003.

\bibitem[Guo et~al.(2016)Guo, Fan, Ai, and Croft]{jguo_sigir16}
Jiafeng Guo, Yixing Fan, Qingyao Ai, and W.~Bruce Croft.
\newblock A deep relevance matching model for ad-hoc retrieval.
\newblock In \emph{Proceedings of the 39th International ACM SIGIR conference}, 2016.

\bibitem[Guo et~al.(2023)Guo, Ton, Liu, and Li]{guo2023inference}
Ruocheng Guo, Jean-Fran{\c{c}}ois Ton, Yang Liu, and Hang Li.
\newblock Inference-time stochastic ranking with risk control.
\newblock \emph{arXiv e-prints}, pages arXiv--2306, 2023.

\bibitem[Järvelin and Kekäläinen(2000)]{Järvelin2000}
Kalervo Järvelin and Jaana Kekäläinen.
\newblock \text{IR} evaluation methods for retrieving highly relevant documents.
\newblock \emph{Proceedings of the 23rd international ACM SIGIR conference}, 2000.

\bibitem[Khattab et~al.(2020)Khattab, Hammoud, and Elsayed]{okhattab_sigir20}
Omar Khattab, Mohammad Hammoud, and Tamer Elsayed.
\newblock Finding the best of both worlds: Faster and more robust top-k document retrieval.
\newblock \emph{Proceedings of the 43rd International ACM SIGIR Conference}, 2020.

\bibitem[Lei et~al.(2015)Lei, Rinaldo, and Wasserman]{lei2015}
Jing Lei, Alessandro Rinaldo, and Larry Wasserman.
\newblock A conformal prediction approach to explore functional data.
\newblock \emph{Annals of Mathematics and Artificial Intelligence}, 2015.

\bibitem[Liu(2009)]{Liu2009LearningTR}
Tie-Yan Liu.
\newblock Learning to rank for information retrieval.
\newblock \emph{Proceedings of the 33rd international ACM SIGIR conference}, 2009.

\bibitem[Papadopoulos et~al.(2002)Papadopoulos, Proedrou, Vovk, and Gammerman]{papadopoulos2002}
Harris Papadopoulos, Kostas Proedrou, Volodya Vovk, and Alex Gammerman.
\newblock Inductive confidence machines for regression.
\newblock In \emph{ECML}, 2002.

\bibitem[Qin and Liu(2013)]{DBLP:journals/corr/QinL13}
Tao Qin and Tie{-}Yan Liu.
\newblock Introducing {LETOR} 4.0 datasets.
\newblock \emph{CoRR}, abs/1306.2597, 2013.
\newblock URL \url{http://arxiv.org/abs/1306.2597}.

\bibitem[Severyn and Moschitti(2015)]{aseveryn_sigir15}
Aliaksei Severyn and Alessandro Moschitti.
\newblock Learning to rank short text pairs with convolutional deep neural networks.
\newblock In \emph{Proceedings of the 38th International ACM SIGIR Conference}, 2015.

\bibitem[Stephen and K.(1976)]{robertson76}
Robertson. Stephen and Jones. K., Sparck.
\newblock Relevance weighting of search terms. journal of the association for information science and technology.
\newblock \emph{27(3):129-146. doi: 10.1002/ASI.4630270302}, 1976.

\bibitem[Vovk et~al.(1999)Vovk, Gammerman, and Saunders]{Vovk1999}
Vladimir Vovk, Alex Gammerman, and Craig Saunders.
\newblock {Machine-learning applications of algorithmic randomness}.
\newblock \emph{Sixteenth International Conference on Machine Learning (ICML-1999)}, 1999.

\bibitem[Vovk et~al.(2005)Vovk, Gammerman, and Shafer]{vovk2005}
Vladimir Vovk, Alexander Gammerman, and Glenn Shafer.
\newblock \emph{Algorithmic learning in a random world}, volume~29.
\newblock Springer, 2005.

\bibitem[Wang and Joachims(2023)]{wang2023uncertainty}
Lequn Wang and Thorsten Joachims.
\newblock Uncertainty quantification for fairness in two-stage recommender systems.
\newblock In \emph{Proceedings of the Sixteenth ACM International Conference on Web Search and Data Mining}, pages 940--948, 2023.

\bibitem[Yin and et~al(2016)]{dyin_kdd16}
Dawei Yin and et~al.
\newblock Ranking relevance in \text{Yahoo} search.
\newblock \emph{Proceedings of the ACM SIGKDD Conference}, 2016.

\bibitem[Yu(2020)]{yu2020ptranking}
Hai-Tao Yu.
\newblock \text{PT-Ranking}: A benchmarking platform for neural learning-to-rank, 2020.

\end{thebibliography}
